\documentclass{article}
\usepackage{graphicx}
\usepackage{amsmath}

\input{tcilatex}
\begin{document}

\title{Expectation Value in Bell$^{\prime }$s Theorem}
\author{Zheng-Chuan Wang \\
Department of Physics, The Graduate School of the Chinese Academy of
Sciences, P. O. Box 4588, Beijing 100049, China.}
\maketitle

\begin{abstract}
We will demonstrate in this paper that Bell$^{\prime }$s theorem (Bell$%
^{\prime }$s inequality) does not really conflict with quantum mechanics,
the controversy between them originates from the different definitions for
the expectation value using the probability distribution in Bell$^{\prime }$%
s inequality and the expectation value in quantum mechanics. We can not use
quantum mechanical expectation value measured in experiments to show the
violation of Bell$^{\prime }$s inequality and then further deny the local
hidden-variables theory. Considering the difference of their expectation
values, a generalized Bell$^{\prime }$s inequality is presented, which is
coincided with the prediction of quantum mechanics.

03.65.Ud, 03.65.Ta.
\end{abstract}

Local hidden-variables theory\cite{bohm} was ever a striking point in the
development of quantum mechanics, yet it$^{\prime }$s validity was
questioned by Bell. In 1964, Bell proposed a famous inequality\cite{bell1}%
(or theorem) and asserted that local hidden-variables theory conflicts with
quantum mechanics and can not reproduce all the prediction of the latter.
This theorem was latterly improved to a new version after Clauser et al.$%
^{\prime }$s experiment\cite{clauser} of testing local hidden-variables
theory, which is suit for the entire family of deterministic and
nondeterministic local hidden-variables theory\cite{bell3}. From then on,
Bell$^{\prime }$s inequality had been discussed widely whatever in theories
or experiments. In 1982, Aspect et al.\cite{aspect} proposed an experiment
to test Bell$^{\prime }$s inequality by use of time-varying analyzers, \
their experimental results coincide with the prediction of quantum mechanics
and indicate Bell$^{\prime }$s inequality is violated. That is a serious
challenge to local hidden-variables theory. However, the debates never
stopped\cite{jackiw}. Jaynes ever criticized that the probabilistic
reasoning in Bell$^{\prime }$s theorem does not follow the rules of
probability theory\cite{jaynes}, Fine discussed the joint distributions and
commtativity in Bell theorem\cite{fine}, there are other viewpoints on Bell$%
^{\prime }$s theorem\cite{green, mermin}, too. Does Bell$^{\prime }$s
theorem really conflict with theory of quantum mechanics? We said:$^{\prime
} $ no.$^{\prime }$ The superficial paradox comes from the confusion of
definition for the expectation value using the probability distribution in
Bell$^{\prime }$s inequality and the expectation value in quantum mechanics.

Consider two particles A and B which have spin 1/2 and are in quantum
mechanical state $|\Psi \rangle _{AB}$. $\mathbf{a}$ and $\mathbf{b}$ are
vectors in ordinary three-space, $A(\mathbf{a,\lambda )}$ is the outcome of
a measurement on $\sigma _{A}\cdot \mathbf{a}$ and $B(\mathbf{b,\lambda )}$
is on $\sigma _{B}\cdot \mathbf{b}$, then the expectation value in Bell$%
^{\prime }$s inequality is 
\begin{equation}
P(\mathbf{a,b)=}\int d\lambda \rho (\lambda )A(\mathbf{a,\lambda )}B(\mathbf{%
b,\lambda ),}
\end{equation}
where $\rho (\lambda )$ is the probability distribution in local
hidden-variables theory. \ We can not identify it with the expectation value 
$_{AB}\langle \Psi |(\sigma _{A}\cdot \mathbf{a)(}\sigma _{B}\cdot \mathbf{b)%
}|\Psi \rangle _{AB}$ in quantum mechanics as the usual proof on Bell$%
^{\prime }$s inequality. In fact, in local hidden-variables theory the
latter can be written as

\begin{equation}
_{AB}\langle \Psi |(\sigma _{A}\cdot \mathbf{a)(}\sigma _{B}\cdot \mathbf{b)}%
|\Psi \rangle _{AB}=\int d\lambda \rho (\lambda )_{AB}\langle \Psi |(\sigma
_{A}\cdot \mathbf{a)(}\sigma _{B}\cdot \mathbf{b)}|\Psi \rangle
_{AB}(\lambda ),
\end{equation}
while the expectation value $P(\mathbf{a,b)}$ in Bell$^{\prime }$s theorem is

\begin{equation}
P(\mathbf{a,b)=}\int d\lambda \rho (\lambda )_{AB}\langle \Psi |\sigma
_{A}\cdot \mathbf{a}|\Psi \rangle _{AB}(\lambda )_{AB}\langle \Psi |\sigma
_{B}\cdot \mathbf{b}|\Psi \rangle _{AB}(\lambda ),
\end{equation}
where we have replaced the outcomes $A(\mathbf{a,\lambda )}$ and $B(\mathbf{%
b,\lambda )}$ of measurements on particles A and B with $_{AB}\langle \Psi
|\sigma _{A}\cdot \mathbf{a}|\Psi \rangle _{AB}(\lambda )$ and $_{AB}\langle
\Psi |\sigma _{B}\cdot \mathbf{b}|\Psi \rangle _{AB}(\lambda )$,
respectively. The discrepancy between (2) and (3) are obvious, especially to
the entangle state. For example, in the case of singlet state, the
expression (3) in Bell$^{\prime }$s theorem is

\begin{eqnarray}
P(\mathbf{a,b)} &=&\int d\lambda \rho (\lambda )\frac{1}{2}[_{A}\langle
\uparrow |\sigma _{A}\cdot \mathbf{a}|\uparrow \rangle _{A}(\lambda
)_{B}\langle \downarrow |\sigma _{B}\cdot \mathbf{b}|\downarrow \rangle
_{B}(\lambda )  \notag \\
+_{A}\langle &\downarrow &|\sigma _{A}\cdot \mathbf{a}|\downarrow \rangle
_{A}(\lambda )_{B}\langle \downarrow |\sigma _{B}\cdot \mathbf{b}|\downarrow
\rangle _{B}(\lambda )  \notag \\
+_{A}\langle &\uparrow &|\sigma _{A}\cdot \mathbf{a}|\uparrow \rangle
_{A}(\lambda )_{B}\langle \uparrow |\sigma _{B}\cdot \mathbf{b}|\uparrow
\rangle _{B}(\lambda ) \\
+_{A}\langle &\downarrow &|\sigma _{A}\cdot \mathbf{a}|\downarrow \rangle
_{A}(\lambda )_{B}\langle \uparrow |\sigma _{B}\cdot \mathbf{b}|\uparrow
\rangle _{B}(\lambda )],  \notag
\end{eqnarray}
and the expectation value (2) in quantum mechanics will become to

\begin{eqnarray}
&&_{AB}\langle \Psi |(\sigma _{A}\cdot \mathbf{a)(}\sigma _{B}\cdot \mathbf{%
b)}|\Psi \rangle _{AB} \\
&=&\int d\lambda \rho (\lambda )\frac{1}{2}[_{A}\langle \uparrow |\sigma
_{A}\cdot \mathbf{a}|\uparrow \rangle _{A}(\lambda )_{B}\langle \downarrow
|\sigma _{B}\cdot \mathbf{b}|\downarrow \rangle _{B}(\lambda )  \notag \\
-_{A}\langle &\uparrow &|\sigma _{A}\cdot \mathbf{a}|\downarrow \rangle
_{A}(\lambda )_{B}\langle \downarrow |\sigma _{B}\cdot \mathbf{b}|\uparrow
\rangle _{B}(\lambda )  \notag \\
-_{A}\langle &\downarrow &|\sigma _{A}\cdot \mathbf{a}|\uparrow \rangle
_{A}(\lambda )_{B}\langle \uparrow |\sigma _{B}\cdot \mathbf{b}|\downarrow
\rangle _{B}(\lambda )  \notag \\
+_{A}\langle &\downarrow &|\sigma _{A}\cdot \mathbf{a}|\downarrow \rangle
_{A}(\lambda )_{B}\langle \uparrow |\sigma _{B}\cdot \mathbf{b}|\uparrow
\rangle _{B}(\lambda )],  \notag
\end{eqnarray}
which clearly show the differences of the expectation values in Bell$%
^{\prime }$s theorem and quantum mechanics. These differences originate from
the correlation between particles A and B.

As a matter of fact, the expectation value in quantum mechanics contains the
correlation between particles A and B, while the expectation value in Bell$%
^{\prime }$s theorem is the product of outcomes of measurements on particles
A and B, the expectation value $A(\mathbf{a,\lambda )}$ of particle A is
independent of the expectation value of particle B and conversely, these
independent outcomes $A(\mathbf{a,\lambda )}$ and $B(\mathbf{b,\lambda )}$
of measurements have destroyed the correlation between particles A and B,
and there are no correlation in the expectation value $P(\mathbf{a,b)}$ of
Bell$^{\prime }$s theorem, which leads to the differences shown in the
above. However, if we chose a non-entangle state $|\Psi \rangle
_{AB}=|\uparrow \rangle _{A}|\downarrow \rangle _{B}$, the differences
between the two expectation values will vanish, because there are no
correlation between A and B in the non-entangle state.

It is this confusion between the two expectation values that leads to the
inconsistency of Bell$^{\prime }$s theorem and quantum mechanics. In the
usual reasoning of Bell$^{\prime }$s inequality, the quantity $\int d\lambda
\rho (\lambda )[A(\mathbf{a,\lambda )}B(\mathbf{b,\lambda )-}A(\mathbf{%
a,\lambda )}B(\mathbf{b}^{\prime }\mathbf{,\lambda )]}$ is naturally divided
as $\int d\lambda \rho (\lambda )\{A(\mathbf{a,\lambda )}B(\mathbf{b,\lambda
)[1\pm }A(\mathbf{a}^{\prime }\mathbf{,\lambda )}B(\mathbf{b}^{\prime }%
\mathbf{,\lambda )]\}}$

$\mathbf{-}$\bigskip $\int d\lambda \rho (\lambda )\{A(\mathbf{a,\lambda )}B(%
\mathbf{b}^{\prime }\mathbf{,\lambda )[1\pm }A(\mathbf{a}^{\prime }\mathbf{%
,\lambda )}B(\mathbf{b,\lambda )]\}}$, where $\mathbf{a}^{\prime }$, $%
\mathbf{b}^{\prime }$ are the other vectors in three -space. This
rearrangement is reasonable in Bell$^{\prime }$s theorem because there are
no correlation between $A(\mathbf{a,\lambda )}$ and $B(\mathbf{b,\lambda )}$
, while errors will occur in quantum mechanics, in which we can not write
out the corresponding expression, i.e. 
\begin{eqnarray}
&&_{AB}\langle \Psi |(\sigma _{A}\cdot \mathbf{a)(}\sigma _{B}\cdot \mathbf{%
b)-}(\sigma _{A}\cdot \mathbf{a)(}\sigma _{B}\cdot \mathbf{b}^{\prime }%
\mathbf{)}|\Psi \rangle _{AB} \\
&=&_{AB}\langle \Psi |(\sigma _{A}\cdot \mathbf{a)(}\sigma _{B}\cdot \mathbf{%
b)}|\Psi \rangle _{AB}[1\pm _{AB}\langle \Psi |(\sigma _{A}\cdot \mathbf{a}%
^{\prime }\mathbf{)(}\sigma _{B}\cdot \mathbf{b}^{\prime }\mathbf{)}|\Psi
\rangle _{AB}]  \notag \\
&&-_{AB}\langle \Psi |(\sigma _{A}\cdot \mathbf{a)(}\sigma _{B}\cdot \mathbf{%
b}^{\prime }\mathbf{)}|\Psi \rangle _{AB}[1\pm _{AB}\langle \Psi |(\sigma
_{A}\cdot \mathbf{a}^{\prime }\mathbf{)(}\sigma _{B}\cdot \mathbf{b)}|\Psi
\rangle _{AB}],  \notag
\end{eqnarray}
which is wrong. We can check it by a simple example. Choosing the angle
between vectors $\mathbf{a}$ and $\mathbf{b}$, $\mathbf{b}$ and $\mathbf{c}$
are $60^{\circ }$ in order and $\mathbf{a}^{\prime }$ and $\mathbf{b}%
^{\prime }$ as the same vector $\mathbf{c}$. According to quantum mechanics,
the left side of expression (6) is $-\frac{1}{2}-\frac{1}{2}=-1$, while the
right side is $-\frac{1}{2}[1\pm (-1)]-\frac{1}{2}[1\pm (-\frac{1}{2})]$,
they are not equal. The above rearrangement is violated in quantum
mechanics, and the derivation of Bell$^{\prime }$s inequality can not be
resumed considering the correlation in quantum mechanics.

So far, we conclude that Bell$^{\prime }$s inequality does not really
conflict with quantum mechanics, the paradox between them originates from
the difference of their expectation values. Bell$^{\prime }$s inequality is
reasonable if we define the expectation value as expression (1) in which the
correlation between particles A and B do not exist. However we can not
simply compare this expectation value with the one in quantum mechanics
because of the above difference. The experimental results did not violated
Bell$^{\prime }$s inequality because it measured the expectation value in
quantum mechanics instead of the expectation value in Bell$^{\prime }$s
inequality. We should not deny the local hidden-variables theory by use of
this kind of experiments. Local hidden-variables theory should be tested by
other ways.

Our analysis have some similarities with Jaynes$^{\prime }$\cite{jaynes},
Jaynes$^{\prime }$ main contention was that Bell$^{\prime }$s factorization
for the probability of joint outcomes A and B of the two measurements does
not follow from the rules of probability theory. Bell$^{\prime }$s
factorization is $P(A,B|\mathbf{a}$, $\mathbf{b,c,\lambda )=}P(A|\mathbf{a}$%
, $\mathbf{c,\lambda )}P(B|\mathbf{b}$, $\mathbf{c,\lambda )}$, while the
correct factorization should be $P(A,B|\mathbf{a}$, $\mathbf{b,c,\lambda )=}%
P(A|B,\mathbf{a}$,$\mathbf{b,c,\lambda )}P(B|\mathbf{a,b}$, $\mathbf{%
c,\lambda )}$, which is analogous to our analysis. In our treatment, the
expectation value containing the correlation between particles A and B can
not be divided into the product of two single expectation values for
particles A and B, that means $_{AB}\langle \Psi |(\sigma _{A}\cdot \mathbf{%
a)(}\sigma _{B}\cdot \mathbf{b)}|\Psi \rangle _{AB}(\lambda )\neq
_{AB}\langle \Psi |\sigma _{A}\cdot \mathbf{a}|\Psi \rangle _{AB}(\lambda
)_{AB}\langle \Psi |\sigma _{B}\cdot \mathbf{b}|\Psi \rangle _{AB}(\lambda )$%
. Although the expectation value $_{AB}\langle \Psi |(\sigma _{A}\cdot 
\mathbf{a)(}\sigma _{B}\cdot \mathbf{b)}|\Psi \rangle _{AB}$ here is
different from the probability in Jaynes$^{\prime }$ reasoning and can not
be factorized into the product of two terms as Jaynes$^{\prime }$, there
still exist similarities between them.

Due to the above analysis, we can not test Bell$^{\prime }$s inequality by
use of the expectation value containing the correlation between two
particles. In Aspect et al.$^{\prime }$s experiment, the expectation value $%
_{AB}\langle \Psi |(\sigma _{A}\cdot \mathbf{a)(}\sigma _{B}\cdot \mathbf{b)}%
|\Psi \rangle _{AB}$ involving the correlation of two photons is measured by
the time-varying analyzers, which is not the same quantity $P(\mathbf{a}$,$%
\mathbf{b)}$ in Bell$^{\prime }$s inequality, we can not conclude that the
experimental results violate Bell$^{\prime }$s inequality, and this kind of
measurements of expectation values can not be used to test Bell$^{\prime }$s
inequality. However, if we measure the expectation values $_{AB}\langle \Psi
|\sigma _{A}\cdot \mathbf{a}|\Psi \rangle _{AB}(\lambda )$ and $_{AB}\langle
\Psi |\sigma _{B}\cdot \mathbf{b}|\Psi \rangle _{AB}(\lambda )$ for \
photons A and B separately, there will no correlation between particles A
and B be involved in the procedure of measurement, the expectation value $%
\int d\lambda \rho (\lambda )A(\mathbf{a,\lambda )}B(\mathbf{b,\lambda )}$
in Bell$^{\prime }$s inequality can thus be obtained naturally, then we can
compare the experimental results with Bell$^{\prime }$s inequality directly,
that will fulfil the test of Bell$^{\prime }$s inequality. We believe that
this kind of experimental results will coincide with Bell$^{\prime }$s
inequality.

Finally, we extend Bell$^{\prime }$s inequality to a general form in order
to match it with the prediction of quantum mechanics. As we know, $\rho
(\lambda )$ is the probability distribution of local hidden-variable $%
\lambda $ in Bell$^{\prime }$s theorem, however, if we interpret it as the
density matrix of quantum state, which is different from the original
definition of $\rho (\lambda )$, we can define the expectation value in Bell$%
^{\prime }$s inequality as

\begin{equation}
P(\mathbf{a,b)=tr[}\int d\lambda \rho (\lambda )(\sigma _{A}\cdot \mathbf{a)(%
}\sigma _{B}\cdot \mathbf{b)],}
\end{equation}
which has the similar form with Bell$^{\prime }$s definition (1), but here
it is identical to the quantum mechanics expectation value (2), and the
difference of expectation values between Bell$^{\prime }$s inequality and
quantum mechanics will vanish. This generalized expectation value (7)
satisfies $-1\leq \mathbf{tr[}\int d\lambda \rho (\lambda )(\sigma _{A}\cdot 
\mathbf{a)(}\sigma _{B}\cdot \mathbf{b)]}\leq 1.$ So we have

\begin{eqnarray}
&&|\mathbf{tr[}\int d\lambda \rho (\lambda )(\sigma _{A}\cdot \mathbf{a)(}%
\sigma _{B}\cdot \mathbf{b)]-tr[}\int d\lambda \rho (\lambda )(\sigma
_{A}\cdot \mathbf{a)(}\sigma _{B}\cdot \mathbf{c)]|} \\
&&\mathbf{+}|\mathbf{tr[}\int d\lambda \rho (\lambda )(\sigma _{A}\cdot 
\mathbf{b)(}\sigma _{B}\cdot \mathbf{c)]|}  \notag \\
&\leq &3,  \notag
\end{eqnarray}
which can be further rearranged as

\begin{equation}
|P(\mathbf{a,b)-}P(\mathbf{a,c)|}\leq 3-|P(\mathbf{b,c)|.}
\end{equation}
So taht we have arrived at our final results- the general Bell$^{\prime }$s
inequality, which has similar form with the original Bell$^{\prime }$s
inequality. Inequality (9) has contained the correlation between particles A
and B, and must coincide with the prediction of quantum mechanics.

In summary, we show that Bell$^{\prime }$s inequality does not conflict with
quantum mechanics because of the difference on the expectation values of
their own. The experimental results have not violated Bell$^{\prime }$s
inequality. Local hidden variables theory still need to be tested by other
ways.

\bigskip $\mathbf{Acknowledgments}$

It is pleasure to acknowledge valuable discussions with Dr. Qing-Rong Zheng,
Biao Jin. This work is supported in part by the NNSF (Grant No. 10404037)
and the fund of Ministry of Science And Technology Grant No. 2002CCA02600.

\end{document}